\documentclass{optica-article}

\journal{opticajournal} 
\articletype{Research Article}

\usepackage{lineno}

\begin{document}

\title{Real time monitoring of pressure-induced deformation of PDMS to evaluate pressure distribution in microfluidic channels}

\author{Kiran Acharya,\authormark{1,2} Serge Monneret,\authormark{1} Martin Brandenbourger,\authormark{2} and Thomas Chaigne\authormark{1,*}}

\address{\authormark{1}Aix Marseille Univ, CNRS, Centrale Med, Institut Fresnel, Marseille, France\\
\authormark{2}Aix-Marseille Univ, CNRS, Centrale Med, IRPHE, UMR 7342, Marseille, France}

\email{\authormark{*}thomas.chaigne@fresnel.fr} 


\begin{abstract*} 
Accurate pressure measurements in micrometric channels are essential for a wide range of microfluidic applications. Existing approaches rely on a variety of sensing mechanisms, but generally require the integration of additional probes or sensing elements during or after chip fabrication. Here, we introduce a pressure sensing approach based on quantitative phase imaging of the deformation of compliant microfluidic channels. We demonstrate real-time measurements of channel deformation over a large field of view with high sensitivity, without the need for embedded components or modifications of the microfluidic device.
\end{abstract*} 


\section{Introduction}
Pressure is a central physical quantity in microfluidic systems. It governs flow rates, determines hydraulic resistance, drives transport through complex networks, and provides a direct diagnostic of device operation \cite{Shenreview2021}. As microfluidic platforms become increasingly integrated and multifunctional, accurate pressure measurement is therefore essential not only for flow control, but also for device characterization, failure detection, and quantitative interpretation of experimental conditions. This is particularly important in lab-on-chip applications involving chemical reactions, fluid mixing, biomedical assays, and organ- or tissue-on-chip systems, where pressure and pressure-induced shear stresses can directly affect living cells and biological samples \cite{convery201930,ingber2022human}. 

Pressure measurement becomes even more critical in deformable microfluidic channels made of soft materials such as PDMS. In these systems, internal pressure induces wall deformation, which in turn modifies the channel cross-section and hydraulic resistance \cite{hardy_deformation_2009,raj_flow-induced_2016}. The resulting fluid-structure interaction can lead to nonlinear flow behavior and pressure-dependent device properties. Consequently, measuring pressure locally inside these systems is not only critical to control flow, but critical to characterize the mechanics of the device itself.

A broad range of pressure measurement methods has been developed for microchannels \cite{Shenreview2021}. These techniques may be distinguished according to several criteria: ability to measure in situ or only outside the chip; point measurement or spatially resolved pressure field; physical sensing mechanism: displacement of a mechanical element, or of a liquid-gas interface, changes in bulk properties (e.g. optical or electrical) of the substrate or the fluid; and whether the sensor requires a dedicated structure integrated into the chip or can operate on an otherwise unmodified device. Important performance metrics include pressure measurement range and sensitivity, robustness to fabrication tolerances and to environmental disturbances, temporal bandwidth, and spatial resolution. 

Most in situ microfluidic pressure sensors rely on a pressure-sensitive element integrated within the chip. In side-channel or trapped-air approaches, pressure is inferred from the displacement of a liquid--gas interface, conceptually similar to a miniaturized manometer. 
In membrane-based sensors, pressure deforms a thin wall or diaphragm, and this deformation can be read out for instance through electrical capacitance or resistance changes. 
A variety of optical approaches have been introduced to provide high sensitivity and compact pressure readout. Some rely on dedicated chip-integrated optofluidic structures whose geometry or optical properties vary with pressure. Membrane displacement can be optically detected, for instance using interferometric methods or confocal microscopy \cite{song_optofluidic_2010,Song2011,Kamruzzaman2019,molla_pressure_2011}. Pressure-induced deformation or refractive-index changes can modify light transmission, color, resonance wavelength, cavity response, fluorescence emission, or speckle patterns \cite{Gaber2019,Li2018,Xiao2018,Ducloue,hardy_deformation_2009,gervais_flow-induced_2006,chung_multiplex_2009,kim_remote_2016}. Related approaches use fiber-optic sensors, including Fabry--Perot cavities, in which pressure changes the cavity length and therefore the reflected or transmitted spectrum \cite{Wu2023}. 

Although these methods can provide highly sensitive \textit{in situ} measurements, they generally require the microfluidic geometry to be specifically designed around the sensing principle. This requirement can increase fabrication complexity, partially fill chip area, perturb the local mechanical or hydraulic properties, and overall limit applicability to already fabricated or generic devices.
Additionally, conventional interferometric implementations can be sensitive to environmental perturbations such as vibrations, air currents, stray reflections, and illumination instabilities. 

Yet optical techniques are particularly attractive for pressure sensing in microfluidics as they provide high sensitivity, non-contact readout, which are compatible with transparent microfluidic materials. With the appropriate imaging configuration, they may also enable spatially resolved measurements over an extended field of view. 
Here, we introduce quantitative optical phase imaging as a fully non-contact method for measuring pressure in simple transparent deformable microfluidic channels. Rather than integrating a dedicated pressure sensor, the method exploits the pressure-induced deformation of the channel itself and reads it out through optical phase variations. By tracking relative channel deformations over the field of view, the technique enables spatially resolved pressure mapping with sub-micrometric displacement sensitivity and temporal resolution compatible with dynamic microfluidic experiments. In the present implementation, the measured deformation sensitivity corresponds to pressure resolutions on the order of a few to a few tens of millibars, depending on the channel geometry and calibration conditions. Because the measurement is self-referenced and image-based, it retains the high sensitivity of interferometric optical methods while reducing sensitivity to environmental vibrations and alignment drifts. This approach therefore provides a simple route toward non-invasive, full-field pressure measurement, with high spatio-temporal resolution (<1 µm, >1Hz), in deformable microfluidic systems without requiring dedicated sensor integration.

\section{Optical phase imaging for pressure measurement in a microfluidic channel} \label{pple}
\subsection{Measurement principle}
When light propagates through a transparent medium with refractive index contrast, the wavefront gets distorted. 
Fig.\ref{newfig1} highlights this phenomenon in a simple microfluidic channel, in which water and the surrounding medium, here polydimethylsiloxane (PDMS), exhibit slightly different refractive indices ($n_{water} = 1.33$ and $n_{PDMS} \simeq 1.4$). For an horizontal incident plane wave, the total time delay for photons to travel through the sample varies with the lateral position in the sample. As a consequence, the shape of the outcoming deformed wavefront is directly linked to the optical path difference (OPD), defined as following for the simple microfluidic channel:
\begin{equation}\label{opd}
\mathrm{OPD}(x) = \int_{y} \left[n_{water} - n_{PDMS} \right] \, dy,
\end{equation}
Considering the specific case of a channel formed by a circular tube of radius R, the theoretical OPD distribution through the chip can be derived from Eq.\eqref{opd}:
\begin{equation}\label{opdcirc}
    \mathrm{OPD}_{\mathrm{th}}(x) = 2 (n_{water} - n_{PDMS}) \sqrt{R^2 - x^2} = -2\,\Delta n\,\sqrt{R^2 - x^2}
\end{equation}
where $\Delta n$ = $n_{PDMS}$ – $n_{water} \simeq 0.07$. 
We neglect here refraction effects at the PDMS-water interface. This holds true around the center of the channel, corresponding to small angles of incidence of light with respect to the channel. However, this assumption becomes less valid at the edges of the channel, which could introduce errors. 

\begin{center}
\includegraphics[width=0.9\linewidth]{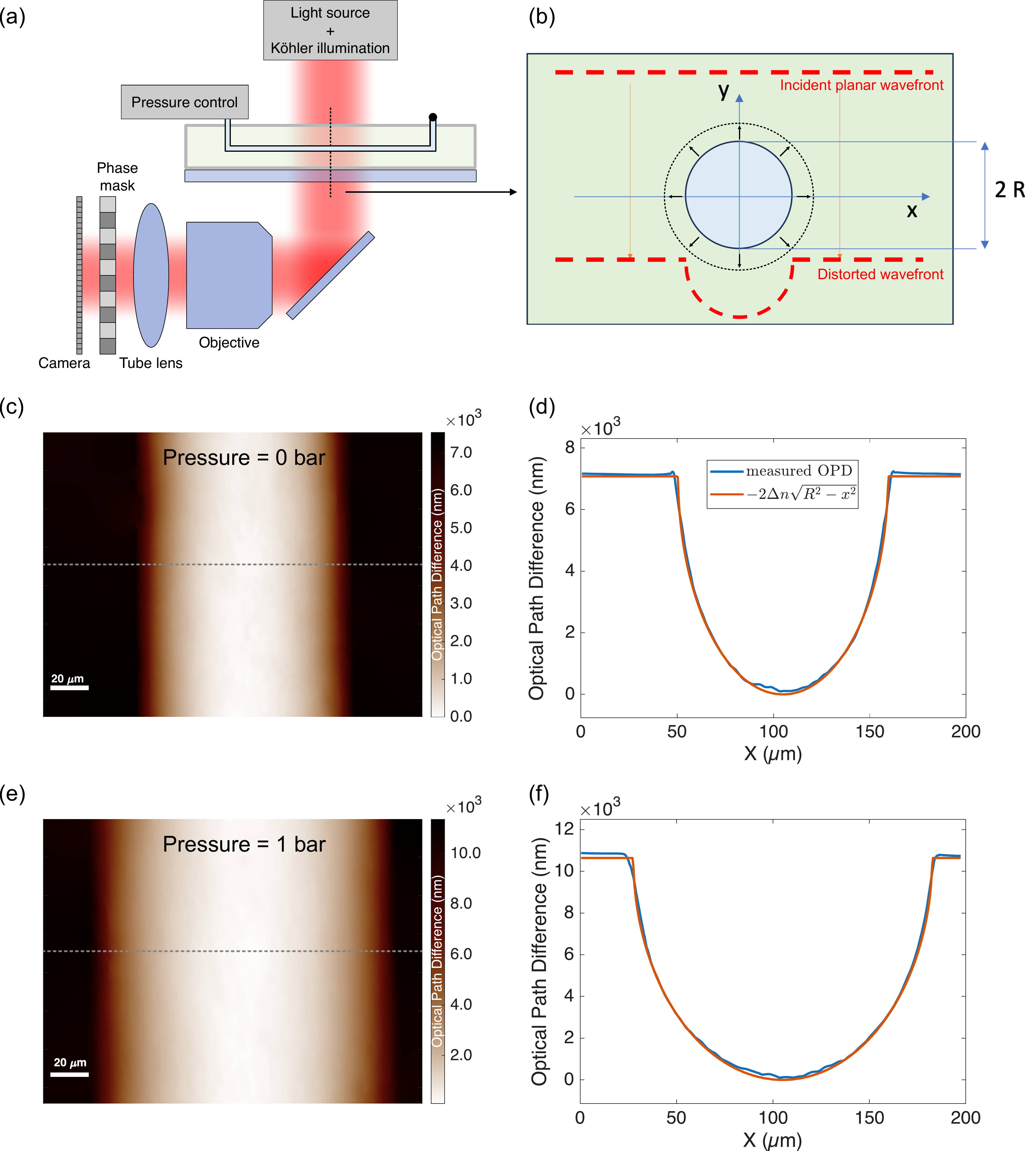}
\captionof{figure}{\textbf{Principle of wavefront analysis through a transparent, deformable microfluidic chip.} (a) Experimental set-up for quantitative phase imaging of a microfluidic chip using a wavefront sensor as the phase analyzer. (b) Schematic of the chip geometry: a linear channel with circular cross-section is embedded in a homogeneous PDMS block. A planer incident wavefront is distorted after propagation through the channel (red dashed lines). (c) Optical path difference (OPD) image of the channel (top view) with no pressure applied, with 700 nm illumination and X60 optical magnification. (d) Experimental transverse OPD profile (along dashed line indicated in c) and fitted model. Fitted parameters are R = 54.5 µm and $\Delta n$ = 0.065. e) OPD image of the channel with 1 bar of pressure applied in the channel. (f) Experimental transverse OPD profile and fitted model. Fitted parameters are R = 77.8 µm and $\Delta n$ = 0.0682.}
\label{newfig1}
\end{center}
\vspace*{10pt}

We map OPD using Quadriwave Lateral Shearing Interferometry (QLSI), to measure the shape of the optical wavefront after its propagation through the chip. 
By placing a two-dimensional binary phase mask in front of a regular camera sensor (Fig.\ref{newfig1}(a)), this OPD distribution can be imaged with high resolution \cite{Primot95,bon2009}. A typical image is shown in Fig\ref{newfig1}(c). 
When light passes through the 2D grating, it is split into four replicas, which will then interfere on the camera sensor. The resulting interferogram exhibits bright spots, which will be spatially shifted by any local tilt of the incoming optical wavefront. The phase gradient, and in turn the OPD, can then be reconstructed from these interferogram local distortions.
Compared to other phase imaging techniques, such a wavefront sensing-based technique can be adapted on any regular optical microscope, by attaching the wavefront analyzer to the camera port. It requires only a single measurement to retrieve phase and intensity, exhibits higher spatial resolution than other wavefront sensing methods, and is achromatic \cite{Chaumet2024,primot_theoretical_2003}.

\subsection{Methods}
\subsubsection{Experimental setup}
The experimental setup we used to realize quantitative phase imaging by wavefront sensing on microfluidic chips is shown in Fig.\ref{newfig1}. It is composed of a conventional inverted microscope (Eclipse Ti2, Nikon, Japan) equipped with a water immersion objective (Nikon CFI Fluor 60x, 1.0 numerical aperture, 2.0 mm working distance) and a transmission Köhler illumination based on an halogen light source, with a reduced aperture diaphragm. The wavefront analyzer (SID4Bio, Phasics SA, Saint Aubin, France) was mounted on a C-mount adapter on one of the microscope’s lateral camera ports so that the detector plane matches the microscope image plane \cite{bon2009}. For pressure control, the chip was connected either to a pressure pump (OB1 MK4, Elveflow), with PTFE tubing (1/16" OD X 1/32" ID).

As QLSI is only sensitive to OPD gradients, no phase pistons can be recorded using the technique we propose, which implies that the only region of the sample contributing to the recorded OPD curve lies within the channel itself (-R<y<+R) if the PDMS is truly homogeneous (see Fig.\ref{newfig1}(c-f)).
During the whole process, the reconstruction algorithm (SID4Bio software, Phasics, France) used to recover the OPD image from the wavefront sensor needs the preliminary recording of a reference image, acquired from an empty part of the sample under the same conditions used for the latter. As such, only the OPD map due to the sample can be displayed, independently of the characteristics of the light beam that may vary according to the conditions under which the optical system is used. In this paper, all reference images were captured near the channel, but filling the field-of-view with only PDMS.

It should also be noted that the native light source of the microscope (white halogen lamp) can be employed to make the images due to the achromatic character of the technique \cite{bon2009}. Nevertheless, when it comes to accurately measuring a refractive index, especially when the material in question exhibits significant chromatic dispersion \cite{Zhang2020}, it is necessary to define a specific spectral band by inserting a bandpass filter into the optical path of the microscope. Unless otherwise noted for certain results presented in this article, we used a filter centered at 700 nm with a bandwidth of ±25 nm for this purpose.

\subsubsection{Channel manufacturing}
As QLSI measures wavefront by integrating phase gradient \cite{Primot95}, typical rectangular channels with steep vertical walls cannot be used. 
The largest wavefront slope that can be measured corresponds to a local shift of the interference fringes over one pixel of the binary phase mask. In our imaging configuration, this corresponds to a change in OPD of about 1 µm over a distance of 0.5 µm. Steeper slopes could be resolved using larger magnification, at the expense of the field-of-view.
Channels with circular cross-section were used instead, to ensure smooth wavefront curvature across the field-of-view. 
The channel was prepared using PDMS as the transparent supporting medium. This elastomer was formed by mixing the PDMS base with a curing agent in a 20:1 ratio to provide high elasticity. For this first proof-of-principle, we specifically sought to work with a highly deformable channel in order to maximize the deformation of the channel for a given pressure.

To produce a smooth circular channel, a PDMS block of 5 mm thickness was molded around a 100 µm diameter nylon wire. Two small vertical PTFE tubes fixed in the plexiglas mold were used to horizontally hold the wire over $\sim 4$ cm, 1 mm away from the PDMS surface. After curing at 80°C for 2h, the PDMS substrate was removed from the mold, and the wire was gently pulled out. This operation leaves a hollow cylindrical channel, with a diameter slightly larger than the wire (see Fig.\ref{newfig1}(c,d)).  
Fig.S1 summarizes the main steps of the manufacturing process. 
The channel was then filled with distilled water, and the inlet was connected to a microfluidic flow controller (OB1 MK4, Elveflow, Paris, France), using stainless steel 90° bent PDMS couplers (Darwin microfluidics). Pressure ramps were applied inside the channel to demonstrate the ability of QLSI to image channel deformation. To maximize this deformation, the outlet was clogged with glue.

\subsubsection{Imaging and data analysis process}
The microfluidic chip was placed in the microscope in order to record interferograms. OPD maps were extracted by the analysis software (see Fig.\ref{newfig1}(c) and (e). A specific Matlab script was developed to extract transverse profiles, and fit them with Eq.\eqref{opdcirc} (see Fig.\ref{newfig1}(d-f)). 
To compare OPD images and compensate for the arbitrary offset introduced in the phase gradient integration step, we set their minimum to zero. 
As mentioned earlier, the simple model given by Eq.\eqref{opdcirc} does not account for refraction at the PDMS-water interface. Considering the large thickness of the channel, this might actually deflect light rays away from the optical axis, and thus generate a non-zero OPD value outside the projection of the channel’s diameter onto the horizontal axis, over a range of a few micrometers. This phenomenon could explain the discrepancy observed in the upper part of the OPD curve. 
Fitted values of R and $\Delta n$ can nevertheless be extracted, provided that the fit properly matches the measured OPD around the center of the channel, where refraction can be neglected.
Consequently, a pressure-induced change in the properties of the channel can be captured from the OPD images, and quantified by fitting these profiles. 

\section{Results}
\subsection{Evolution of the channel geometry with pressure}
To demonstrate the ability of QLSI to measure channel deformations, a pressure ramp was applied in the channel using the pressure pump, ranging from 0 to 1000 mbar with 100 mbar steps. 
At each step, a full OPD image was acquired over the entire channel. 
Fig.\ref{newfig2}(a) shows the evolution of OPD over a 13 µm region around the dashed line in Fig.\ref{newfig1}(c). These OPD images clearly highlight the channel diameter increase with pressure. The maximum OPD value itself, defined as the difference between the outside of the channel and the center of the channel, is also increasing, as predicted by Eq.\eqref{opdcirc}. OPD image profiles in Fig.\ref{newfig2}(b) highlights these effects as well.
As shown in Fig.\ref{newfig1}(e-f), experimental data and model properly match up to 1 bar, validating the proposed model.
In particular, this means that an homogeneous refractive index within the PDMS can still be assumed at this high pressure. 
OPD profiles were fitted to extract channel radius R and refractive index contrast $\Delta n$ as a function of pressure.
Fig.\ref{newfig2}(c) shows the evolution of the channel radius as a function of pressure. 
We observe a significant strain of more than 40\% at maximum pressure, and the experimental curves clearly reveal non-linearities in the deformation. 

\begin{center}
\includegraphics[width=0.9\linewidth]{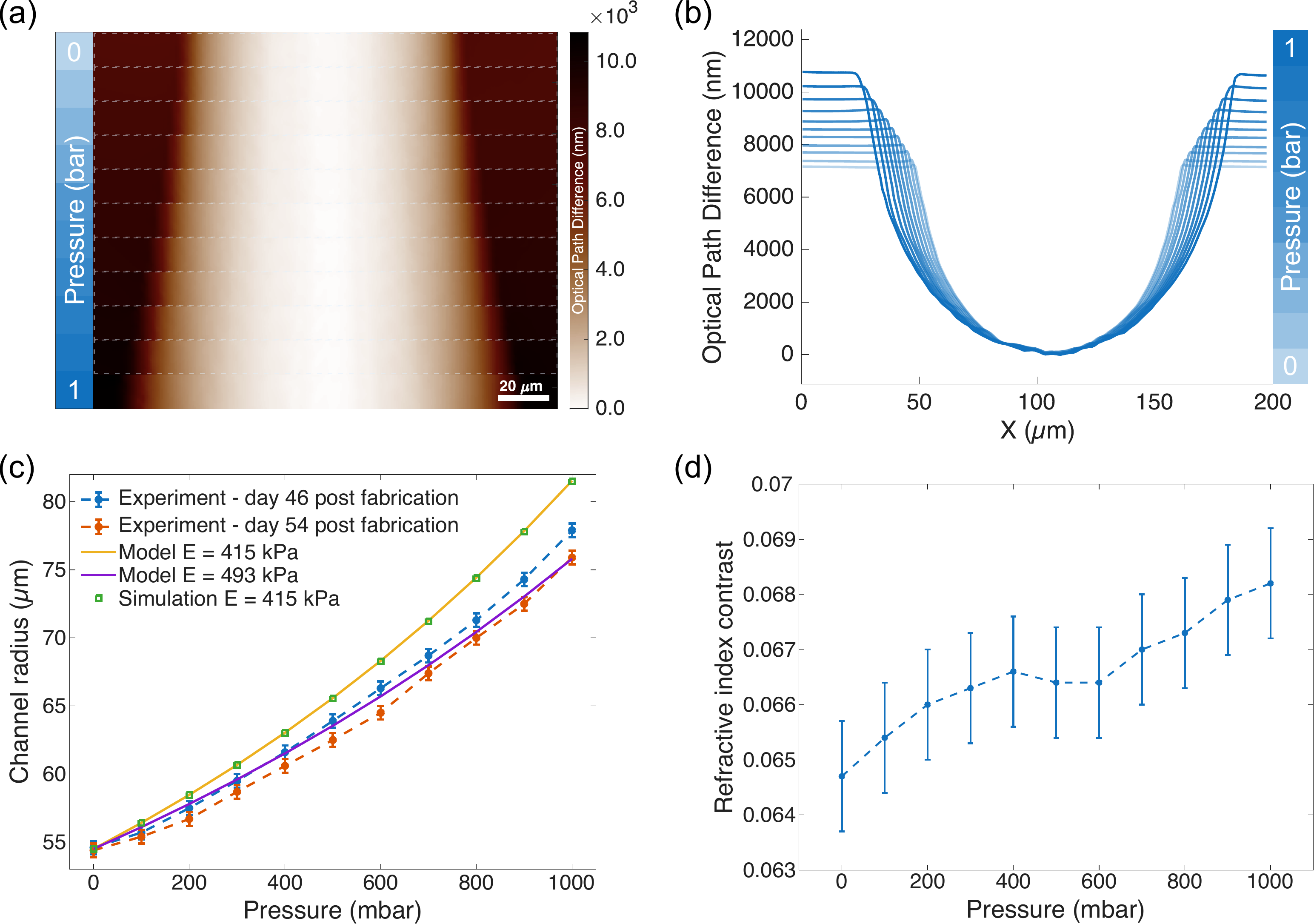}
\captionof{figure}{\textbf{Quantitative phase imaging can measure pressure from microfluidic channel deformation.} (a) 13 µm-high slices (around dashed line indicated in Fig.\ref{newfig1}(c)) of consecutive OPD images under increasing pressure are stacked to highlight channel deformation as a function of pressure. (b) Experimental transverse OPD profiles under increasing pressure from 0 to 1 bar. (c) Evolution of channel radius R as a function of pressure. Dashed lines show fitted values from experimental OPD profiles. Plain lines show theoretical values from Neo-Hookean hyperelastic deformation model. Young moduli of 415 kPa (yellow) or 500 kPa (purple) are compared, for a Poisson's ratio of 0.41. Squares: simulated values from a finite-element model of the deformation (Young modulus: 415 kPa, Poisson's ratio: 0.41). (d) Evolution of refractive index contrast as a function of pressure. All errors bars were assessed as described in section \ref{accu}.}
\label{newfig2}
\end{center}
\vspace*{10pt}

To better grasp these effects, we computed the expected deformation using an hyperelastic Neo-Hookean model \cite{rivlin_large_1948,zulkifli_comprehensive_2023}. The specific 20:1 PDMS chip that was used in these experiments was mechanically characterized through a tensile test (Zwick-Roell zwickiLine Z0.5 TS). Following curing, part of the PDMS sample was cut into small pieces (2.29 mm x 4.47 mm cross-section), and tension force was applied to pull it up to 2 mm and then released back to initial position. 4 experiments of 3 cycles each were performed, yielding an average Young's modulus of $E = 415 kPa$. Poisson's ratio was set as $\nu = 0.41$ \cite{Smith2024}.
Using these parameters, the non-linear Neo-Hookean model slightly overestimates the radius increase with pressure (Fig.\ref{newfig2}(c), yellow line). A potential explanation is that the experiments were performed several days later than the mechanical characterization. The two experiments (Fig.\ref{newfig2}(c), blue and orange lines and dots) presented were themselves performed 8 days apart on the same microfluidic chip, and exhibit slightly different deformations. This difference can be attributed to PDMS stiffening over time. 
Empirically, a Young's modulus of 493 kPa would better explain the observed channel radius increase (Fig.\ref{newfig2}(c), purple line). This would correspond to a increase of Young's modulus of approximately 1.5 kPa/day, in agreement with previously reported stiffening of similar material \cite{zhang_effects_2025}.

To study whether the channel deformation could also be affected by the specific geometry of the chip compared to the infinite PDMS medium assumed in the non-linear model, a two-dimensional finite element model of the microchannel cross-section was developed in COMSOL Multiphysics. The geometry consisted of a rectangular PDMS domain (horizontal 3 cm ×  vertical 5 mm) containing a circular cavity of radius 55 µm representing the water-filled channel. This circle was horizontally centered and placed 1 mm away from the bottom boundary, where the glass substrate is located in the real chip.
A Young’s modulus of $4.15 \times 10^{5} Pa$, a Poisson’s ratio of 0.41, and a density of $970 kg/m^3$ were assigned to the PDMS domain. The fluid domain was modeled as water with a density of $1000 kg/m^3$ and a dynamic viscosity of $10^{-3} Pa·s$.
The Solid Mechanics module was applied to the PDMS region using a hyperelastic Neo-Hookean constitutive model in order to account for the large deformations. To reproduce the experimental boundary conditions, the lower boundary of the PDMS block was fixed to mimic bonding to the rigid glass substrate, while all other external boundaries were left mechanically free, allowing unconstrained lateral and vertical deformation. The computational domain was discretized using quadrilateral finite elements with a maximum element size of 0.35 µm. In addition, a boundary-layer mesh refinement was introduced at the PDMS–water interface to improve the accuracy of deformation and stress calculations near the channel wall.
As can be seen in Fig.\ref{newfig2}(c) (green squares), the effect of the finite distance between channel and substrate barely affects the deformation, reducing only by 0.6 µm the channel radius at 1 bar.

\subsubsection{Evolution of refractive index contrast $\Delta n$ with pressure}

As reported in literature, PDMS 20:1 is a compressible material \cite{Smith2024}. This indicates that the channel expansion under water pressure is accompanied by a change in the density of the PDMS around the channel, and thus in its refractive index. 
Fig.\ref{newfig2}(d) shows that the refractive index constrast $n_{PDMS} - n_{water}$ increases with pressure, with an average slope of $\sim 3 \cdot 10^{-3}$ per bar. 
Pressure dependence of these refractive indices has been already reported in the literature. Regarding water, the change is on the order of $10^{-5}$ per bar \cite{Waxler1963}. We can therefore attribute most of the observed change in $\Delta n$ to a change in the refractive index of PDMS itself, which undergoes compression as the pressure increases in the channel. Previous studies report changes in refractive index of PDMS from $10^{-4}$ per bar to $10^{-3}$ per bar \cite{Park2018,tarjanyi_effect_2014,Turek2014}. The difference can be explained by the PDMS formulation, as well as by the geometry of the sample and the pressure distribution, allowing for different distributions of locally imposed stresses and strains. As the refractive index increase is driven by the density change, this should therefore be normalized by strain rather then pressure, as done in \cite{tarjanyi_effect_2014}, which established comparable refractive index change at similar strain amplitude. Additionally, Turek and coworkers \cite{Turek2014} showed that the change in the refractive index of PDMS increases with longitudinal strain, and that this increase depends significantly on Poisson's ratio. In particular, for a longitudinal strain of 20\%, they report a change in the PDMS refractive index of 0.01 for $\nu$ = 0.45 and 0.025 for $\nu$ = 0.40. Such a variation in proportion to the coefficient (1 - 2 $\nu$) is also predicted by theory, which states that the refractive index change induced by deformation can be derived from the Lorentz–Lorenz relation assuming constant molecular polarizability and relating density change to volumetric strain \cite{Born_Wolf_1999}.

As the edges of the model from Eq.\eqref{opdcirc} remain consistent with the measured OPD profile even at high pressure, we infer that compression induces a refractive index change occurring over a spatial scale larger than the field of view. Finite-element simulations confirm that the radial stress decrease around the channel over a distance of about 50 µm outside the channel (see Fig.S2).

Although not strictly required to monitor pressure, this refractive index change could still be used to further characterize locally the substrate of deformable microfluidic devices.

\subsection{Estimating measurement accuracy} \label{accu}
When fitting the model from Eq.\eqref{opdcirc}, each of the two parameters R and $\Delta n$ have a specific influence on the shape of the OPD profile.
Fig.S3 shows how this shape changes as R and $\Delta n$ vary around their nominal values (at zero pressure). This figure highlights the importance of the upper portion of the measured profiles, that is at the edges of the channel, which represents the region where the fitting procedure is most sensitive.

\begin{center}
\includegraphics[width=0.9\linewidth]{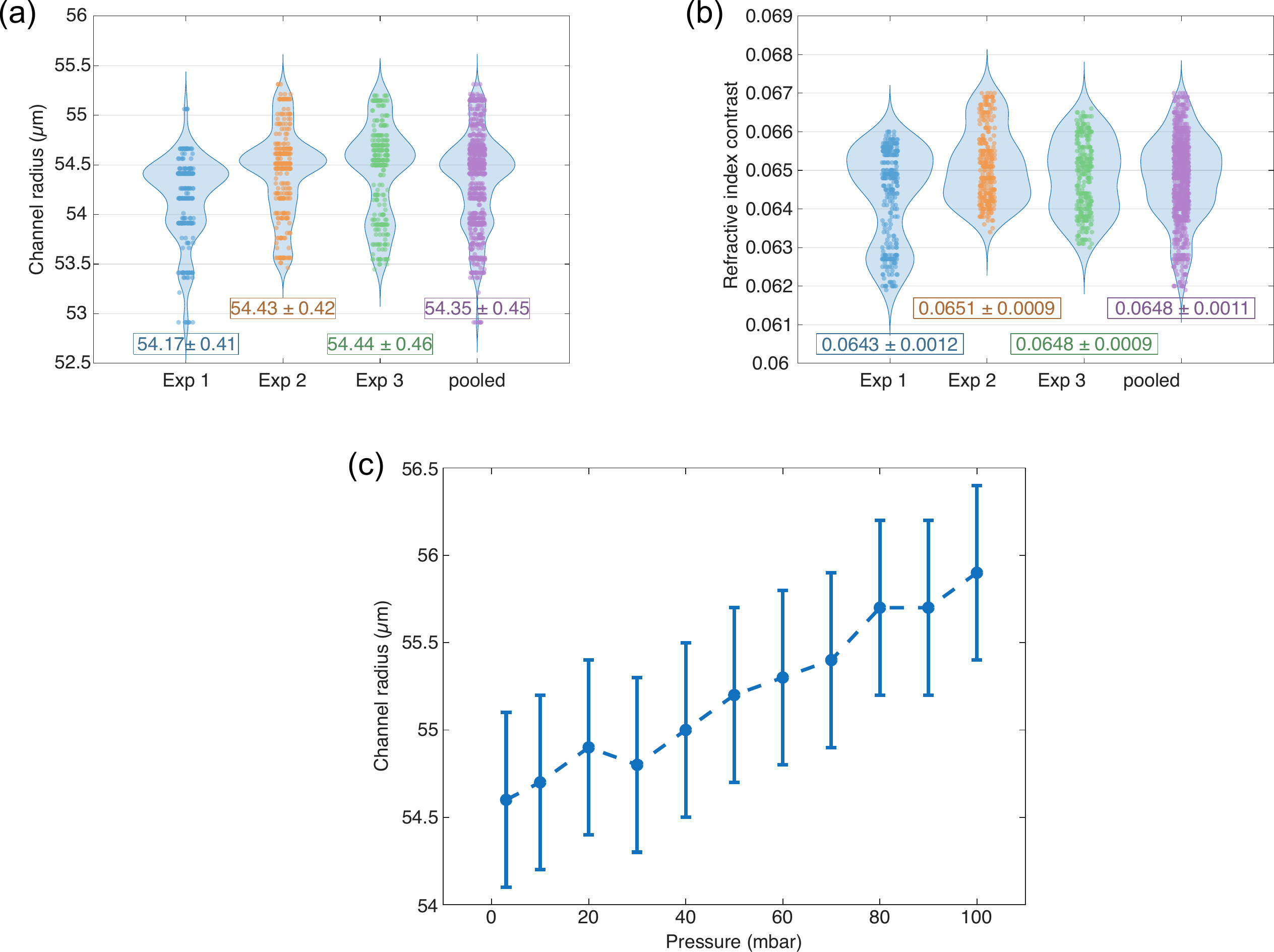}
\captionof{figure}{\textbf{Deformation measurement accuracy and sensitivity.} (a-b) Distribution of R (a) and $\Delta n$ (b) parameters when fitting each line from a single OPD image, for 3 independent experiments with identical experimental configuration (60x magnification, no pressure). (c) Evolution of channel radius for a pressure ramp from 0 to 100 mbar, with step of 10 mbar, highlighting the sensitivity of our technique.}
\label{newfig3}
\end{center}
\vspace*{10pt}

However, the channel edges are also the regions that are most sensitive to focusing during imaging, making comparisons between independent experiments difficult. To evaluate the accuracy of our measurements, we acquired three fully independent phase images, each obtained after removing and reinserting the channel into the microscope. This procedure accounts for the various experimental factors that may affect the measurements, and therefore enables estimation of the overall uncertainty on both the channel radius and the refractive index contrast.

Figure \ref{newfig3}(a,b) shows the results of the analysis. For each image, we analyzed all 300 lines and independently fitted them to evaluate R and $\Delta n$, showing that R was distributed over a range of approximately ±0.5 µm, whereas $\Delta n$ was distributed over ±0.001. 
The mean values of R and $\Delta n$ obtained for different experiments all fall within this range.
This shows a smaller inter-experiment variability explained by the reliability of the method, compared to the intra-image variability mostly coming from a spatial heterogeneity of the PDMS sample. For this reason, in relevant experiments designed to monitor the pressure in the channel, we will select an optimal region in the image that exhibits as uniform a profile as possible, as shown in Fig.\ref{newfig1}.

Based on these results, we estimate that the error bars associated with the measurement of the $\Delta n$ and R values are ±0.001 and ±0.5 µm, respectively. These error bars are displayed in Fig.\ref{newfig2}(c-d), and Fig.\ref{newfig3}(c).
For channel radius R, this accuracy is similar to the size of a pixel in the object plane (29.6 µm in the image plane, considering a 60x magnification). One could therefore argue that similar accuracy could be obtained by measuring the channel deformation from regular intensity images. However, this would only enable to detect the horizontal deformation, while our technique can retrieve the whole shape of the channel. Moreover, diffraction fringes at the edges strongly impair the accuracy of intensity-based measurements (see Fig.S4). These two effects combined would also drastically increase the measurement sensitivity to focusing errors.

Considering the relationship between pressure and channel radius as established in Fig.\ref{newfig2}(c), showing a radius change of 23.3 µm for 1 bar of pressure, the pressure accuracy corresponding to this 0.5 µm error on the channel radius is approximately $0.5*1000/23.3 \simeq 21.46$ mbar. This does not account for non-linearity, the slope and therefore the sensitivity being smaller at low pressure.
To further test the sensitivity of our technique, we performed another experiment in which we increased pressure from 0 to 100 mbar with steps of 10 mbar. Results are shown in Fig.\ref{newfig3}(c).

Whereas results shown in Fig.\ref{newfig3}(a,b) primarily quantify the accuracy of the absolute radius measurement, the sensitivity appears to be higher when considering the pressure-induced radius changes in one specific imaging configuration. This can be estimated from the root-mean-square deviation of the measurements with respect to a linear fit. Using this approach, we obtain a sensitivity of approximately 0.07 µm on the channel radius, smaller than the uncertainty on the absolute radius measurement. 
Considering the lower slope at low pressure, this corresponds to an equivalent pressure sensitivity of 5 mbar.

\subsection{White-light pressure monitoring} 
As previously mentioned, the wavefront sensor used for OPD measurements relies on an achromatic interferometric technique, enabling measurements to be performed with a broadband white-light source. However, for samples exhibiting chromatic dispersion, such as PDMS, it is preferable to perform experiments using spectrally filtered illumination in order to accurately interpret the OPD and be able to convert it to physical thickness using a well defined refractive index.

Nevertheless, thanks to the achromatic nature of the method, the shape of the OPD curve obtained with white light should be similar to that obtained with spectrally filtered light. 
To test this, we imaged the channel with increasing pressure (0, 500, 1000 mbar) using white light and fitted the resulting profiles using the same model from Eq.\eqref{opdcirc} to extract values of R and $\Delta n$. This last parameter can be here interpreted as the average refractive index difference between water and PDMS over the total spectral bandwidth of the halogen white light source (\~ 500-900nm). Results are presented in Fig.\ref{newfigWhite}. We confirm that Eq.\eqref{opdcirc} can equally well fit the OPD profiles from white light measurements, yielding similar value for R.

\begin{center}
\includegraphics[width=0.9\linewidth]{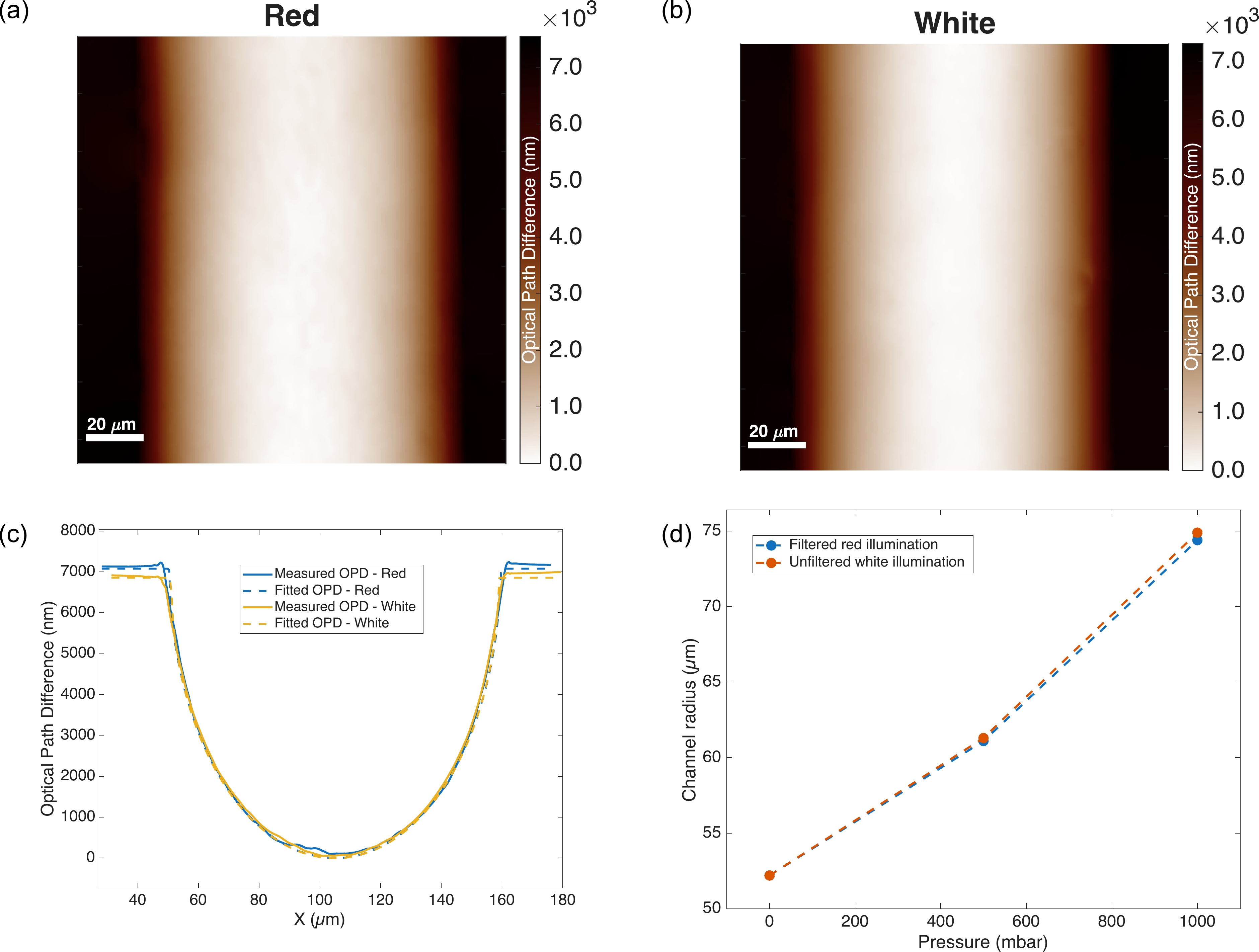}
\captionof{figure}{\textbf{Pressure measurement with white light illumination.} (a) OPD image using filtered red illumination around 700 nm. (b) OPD image using unfiltered white illumination (no pressure applied). (c) Measured and fitted OPD profiles, from OPD image under red or white light. Fitted parameters for white light are R = 54.4 µm and $\Delta n$ = 0.063. The wavelength-dependent optical refractive index of PDMS is seen from the side values in the profiles, highlighting the effect seen in Fig.\ref{newfig3}. (d) Measurement of channel radius for 3 pressure values (0, 500 mbar and 1 bar), under filtered red (orange) or white light (blue) illumination.}
\label{newfigWhite}
\end{center}
\vspace*{10pt}

Pressure monitoring can therefore be performed using the full spectral bandwidth of the source by tracking only the channel radius. This approach benefits from the increased light throughput available under white-light illumination, leading to improved signal-to-noise ratio and-or reduced exposure times. It therefore offers a simpler experimental implementation while also enabling high-speed measurements of dynamic events.

\subsection{Monitoring of dynamic pressure steps}
To investigate this capability, we generated rising or falling pressure steps of 500 mbar and recorded a sequence of OPD images at a 2 Hz frame rate, enabled by the reduced acquisition time due to white light illumination. Fig.\ref{newfigTemp} shows how the channel radius reacts to these sudden pressure steps, thereby providing access to the dynamics of the channel deformation.

\begin{center}
\includegraphics[width=1\linewidth]{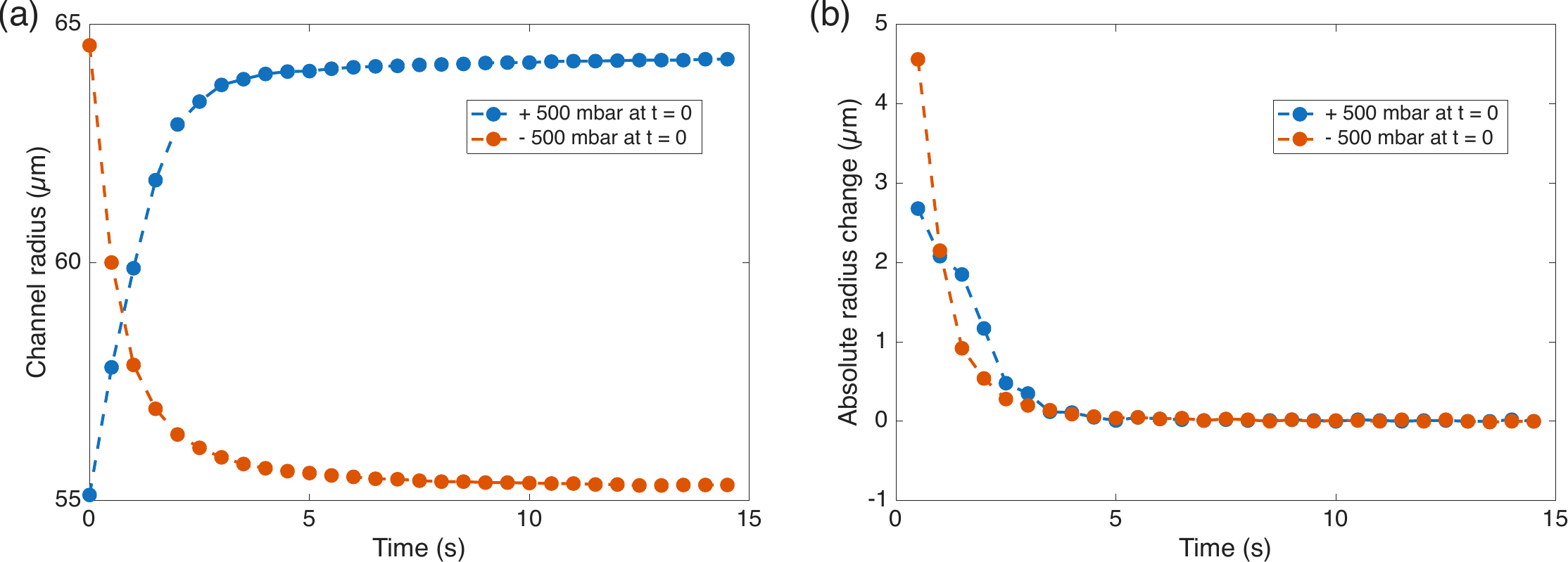}
\captionof{figure}{\textbf{High temporal resolution of channel deformation.} a) Temporal evolution of channel radius upon large 500 mbar pressure step (positive or negative). b) Absolute value of the derivative of the channel radius, highlighting the asymmetry of deformation dynamics between rising or falling pressure.}
\label{newfigTemp}
\end{center}
\vspace*{10pt}

These results also showcase the reliability of the method to study dynamic pressure, as it clearly highlight that no significant drift occurs between consecutive measurements, and that the relative precision is indeed below 0.1 µm for the channel radius.

\section{Discussion}
In this work, we demonstrate that pressure variations inside a transparent deformable microfluidic channel can be measured optically by imaging the pressure-induced deformation of the channel. The approach relies on quantitative phase imaging to measure optical path difference changes associated with the deformation of the channel wall and surrounding PDMS. These measurements are sufficiently sensitive and stable to resolve sub-micrometric, pressure-dependent shape variations, and the measured deformation profiles are well described by an analytical mechanical model. This confirms that, after calibration, optical phase imaging can provide a non-contact route to pressure monitoring in soft microfluidic channels.

A central outcome of this study is that the measured optical signal cannot be interpreted as pressure without an appropriate mechanical calibration. The conversion from optical path difference to pressure depends on the geometry of the channel and on the mechanical properties of the surrounding elastomer. In particular, our comparison between experiments performed at different times highlights that the calibration is age-dependent. This is consistent with the progressive stiffening of PDMS over time, which modifies the amplitude of pressure-induced deformation. As a consequence, a calibration performed shortly after chip fabrication cannot necessarily be used quantitatively several days later. For accurate pressure measurements, the pressure-to-deformation calibration should therefore be repeated regularly, or complemented by an independent characterization of the time-dependent stiffness of each chip.

This sensitivity to material properties is both a limitation and an opportunity. On the one hand, it means that the method is sensitive to several experimental factors, including PDMS aging, fabrication variability, bonding conditions, channel geometry, local inhomogeneities, and possible residual stresses. These parameters can affect the deformation profile and therefore the inferred pressure. On the other hand, the same sensitivity makes quantitative phase imaging a useful tool for characterizing the opto-mechanical properties of transparent soft materials, beyond the sole pressure measurement. In addition to pressure-induced deformation, our measurements provide access to quantities such as the absolute refractive index of PDMS and its variation under mechanical stress. The values obtained are in good agreement with previous reports, supporting the consistency of the optical and mechanical interpretation. This is of particular importance as the specific optical properties of this elastomer strongly depend on the actual fabrication parameters \cite{cai_new_2013,cruz-felix_pdms_2019,Zimmermann2025}. Our technique could therefore be used to accurately characterize these \textit{in situ}.

The present demonstration was performed on a circular cross-section channel, for which the smooth height profile produces a continuous and well-defined optical path difference gradient. However, a circular geometry is not a strict requirement. More generally, the method requires a sufficiently smooth and monotonic optical path difference profile so that small pressure-induced changes can be robustly detected and fitted. For example, a rectangular channel rotated by \(\pi/4\), or other geometries producing smooth height gradients along the imaging direction, could also be suitable. Conversely, abrupt edges, irregular profiles, misalignment with respect to the imaging axis, surface defects, or local inhomogeneities can impair the fitting procedure and reduce the accuracy of pressure estimation. Careful choice of the image profiles used for analysis is therefore essential.
We used channels with a diameter of approximately \(100~\mu\mathrm{m}\), which still exhibit steep OPD  gradients on the edges. A high magnification was therefore required to prevent any artefact in the OPD reconstruction, thus limiting the field-of-view. Smaller cylindrical channels would enable to image over a larger field-of-view, which would be better suited to investigate complex networks.
Larger fields of view or dense microfluidic networks introduce the usual trade-off between spatial resolution and field of view. This trade-off will be important for future applications aiming at pressure mapping over extended microfluidic systems.

The main long-term interest of the method lies in its potential for wide-field pressure mapping. Unlike point sensors, quantitative phase imaging can monitor deformation at many positions simultaneously within the field of view. This opens the possibility of tracking pressure variations at multiple locations in a microfluidic network without integrating dedicated sensors at each point. Recent developments in high-resolution wavefront sensors may further extend this capability by increasing the accessible field of view while preserving phase sensitivity \cite{Wattellier2024}. In this perspective, the technique could become a useful diagnostic tool for complex soft microfluidic systems, where local pressure, deformation, and hydraulic resistance are coupled.

More broadly, the approach is not restricted to pressure sensing in microchannels. It could be extended to other situations where the deformation of a transparent soft material carries useful mechanical information. Examples include monitoring swelling, stress-induced deformation, soft actuators, deformable optical elements, or mechanically responsive polymer structures. In all cases, the main requirement is that the deformation produces a measurable and interpretable optical path difference change.

\section{Conclusion}
We have shown that quantitative optical phase imaging can detect pressure-induced deformation in transparent PDMS microfluidic channels with high sensitivity. The measured deformation profiles are consistent with an analytical mechanical model, enabling pressure estimation after calibration. The method is fully optical, non-contact, and compatible with imaging over an extended field of view, making it attractive for future wide-field pressure mapping in deformable microfluidic systems. These results open the way to more accurate characterization of nonlinear processes in microfluidic devices \cite{keiser2022intermittent,battat2022nonlinear,case2020spontaneous}, and provide a route to explore microscale nonlinear flows previously restricted to larger scales or theoretical descriptions\cite{winn2026unidirectional,garcia2025spontaneous,gomez2017passive,brandenbourger2020tunable}.

The main limitation is that quantitative pressure estimation depends strongly on calibration. In particular, PDMS stiffening over time modifies the pressure-to-deformation response, requiring regular calibration or independent monitoring of the chip mechanical properties. The method is also sensitive to channel geometry, alignment, and local fabrication inhomogeneities, which must be carefully controlled to obtain reliable deformation profiles.

Despite these limitations, the results demonstrate a promising route toward non-invasive pressure monitoring in soft transparent microfluidic devices. By combining optical phase sensitivity with mechanical modeling, the method provides both a pressure-sensing strategy and a tool for characterizing the opto-mechanical response of deformable polymer microchannels.\\
\\
See Supplement 1 for supporting content.\\
\\
\textbf{Funding.} KA acknowledges funding from Aix-Marseille University. MB acknowledges funding from the European Research Council
under Grant Agreement No. 101117080. TC acknowledges funding from the European Research Council
under Grant Agreement No. 101117471.\\
\\
\textbf{Disclosures.} The authors declare no conflicts of interest.\\
\\
\textbf{Data availability.} Data underlying the results presented in this paper are not publicly available at this time but may be obtained from the authors upon reasonable request.


\bibliography{Biblio}

\end{document}